\documentclass[11pt]{article}
\usepackage{amsfonts,amssymb,latexsym,amsthm}
\usepackage{amsmath,amscd}
\usepackage{amsmath}
\usepackage{graphicx}
\usepackage{epstopdf}
\usepackage{subfigure}
\usepackage{mathrsfs}
\usepackage{url}
\newcommand\dif{\mathrm{d}}
\makeatletter

\newtheorem{lemma}{Lemma}
\newtheorem{remark}{Remark}
\newtheorem{proposition}{Proposition}
\newtheorem{theorem}{Theorem}
\newtheorem{corollary}{Corollary}
\newtheorem{conjecture}{Conjecture}
\begin{document}
\begin{flushleft}
{\Large
\textbf{The network-level reproduction number and extinction threshold for vector-borne diseases  }
}
\\
Ling Xue $^{\ast}$,
Caterina Scoglio $^{}$
\\
\bf Kansas State Epicenter, Department of Electrical \& Computer Engineering, Kansas State University, \ U.S. \ 66506
\\
$\ast$ E-mail: lxue@ksu.edu
\end{flushleft}

\section*{Abstract}
The reproduction number of deterministic models is an essential quantity to predict whether an epidemic will spread or die out.  Thresholds for disease extinction contribute  crucial  knowledge on disease control, elimination, and mitigation of infectious diseases. Relationships  between the basic reproduction numbers  of two network-based   ordinary differential equation vector-host models, and extinction thresholds of corresponding continuous-time Markov chain models are derived under some assumptions.  Numerical simulation results for malaria and Rift Valley fever transmission on  heterogeneous networks are in agreement with analytical results without any assumptions,  reinforcing the relationships may always exist and proposing a mathematical problem of proving their existences in general. Moreover, numerical simulations show that  the reproduction number is not monotonically  increasing or decreasing with the extinction threshold.   Key parameters in predicting uncertainty of extinction thresholds   are identified using Latin Hypercube Sampling/Partial Rank Correlation Coefficient. Consistent trends of extinction probability observed through numerical simulations provide novel insights into mitigation strategies to increase the disease extinction probability.  Research findings may improve understandings  of  thresholds for disease persistence in order to control vector-borne diseases.

Keywords:  Extinction thresholds; Reproduction number; Network; Branching process; Vector-borne diseases
\section{Introduction}
Vector-borne diseases  greatly impact health of humans and animals and are among the leading causes of worldwide death every year  \cite{Gratz1999}; almost half of the world's population is infected with at least one type of vector-borne diseases and millions of people die of vector-borne diseases every year \cite{CIESIN2007}. These diseases  also cause significant economic losses in areas of animal trade, agriculture,  health care,  tourism, as well as destroy ecosystems of throughout the world.  Therefore, control and prevention of vector-borne diseases are both economical and humane. Efficient  interventions require a good understanding of  disease transmission and persistence, and dynamic modeling of vector-borne diseases may contribute greatly to this end \cite{Finkenstadt2002}.  A model may be used to learn many characteristics of an outbreak such as: whether or not an outbreak may occur, the size of the outbreak, the duration time of the outbreak, or the probability for the epidemic to die out \cite{Britton2009}. Efficient mitigation strategies deduced from model results may stop an outbreak at early stages by reducing spreading parameters \cite{Britton2009}.

Globalization of trade and travel is one of the key  factors  driving the emergence of vector-borne diseases; heterogeneous structure also plays an important role on dynamics of infectious diseases \cite{kao2010}.  Modeling the spatial spread of vector-borne diseases is a challenging task \cite{Arino2003}, but one possible approach is to consider a meta-population  as a directed graph, or a network, with each vertex representing a subpopulation in  a location, and links placed between two locations if there is a possibility of transmission, such as movement or proximity \cite{Bisanzio2012}.  Network models are more  widely used  in epidemiology to understand the spread of infectious diseases through  connected populations \cite{Natale2006,Vernon2009}.

The basic reproduction number, $R_0$ defined as the number of secondary cases produced by an infected individual in a naive population \cite{diekmann1990definition} is an important threshold  on epidemiology among many others, such as type reproduction number \cite{Heesterbeek2003}, target reproduction \cite{Shuai2012}, and threshold index for epidemicity \cite{Hosack2008}. The basic reproduction number is an important metric, predicting whether a disease will spread or die out in deterministic population and communicable disease theory \cite{Lindarelation2013}. If $R_0>1$, one infectious individual  generally produces more than one infection, leading to spread of an epidemic, whereas if $R_0<1$, one infectious individual generates less than one infection on average \cite{Hosack2009}, and epidemic may die out \cite{diekmann1990definition}. The same trajectory can always be observed with deterministic models given the same initial conditions \cite{Keeling2008}.  If it is possible for an  epidemic to occur again, a real world epidemic does not allow us to observe that the same infection happens to the same person at the same time \cite{Keeling2008}. Moreover, deterministic models have
the shortcoming that the number of infected individuals  may go to  less than one \cite{Lloyd2007}.

	In comparison,  Markov chain models  are more realistic in the sense of only taking integer values instead of continuously varying quantities \cite{Lloyd2007} and taking into account chances by approximating or mimicking the random or probabilistic factors.
The last infectious individual may recover before the infection is transmitted to other susceptible
individuals so that the disease may become extinct \cite{Lloyd2007}. Consequently, an infection introduced to a completely susceptible population  may not invade the system even if $R_0>1$ \cite{Lloyd2007}. Threshold  for the  extinction of an infectious disease to occur and probability of disease extinction are of interests. Bienaym\'{e}-Galton-Watson   branching processes are widely used to study  extinction of diseases involving multi-type infections.



Lloyd  \cite{Lloyd2007} reviewed theory of branching processes  and computed extinction probability using branching processes for  Ross malaria model \cite{Ross1911} taking into account stochasticity and heterogeneity.  P\'{e}nisson   \cite{Penisson2010} presented several statistical tools to study  extinction of populations consisted of different types of individuals, and their behaviors before extinction and in the case of a very late extinction. Allen and Lahodny  Jr \cite{Allen2012}  computed   reproduction numbers for deterministic models, and extinction thresholds for corresponding continuous-time Markov chain (CTMC) models using continuous-time branching process,  and derived their relationships.  A CTMC  model is a stochastic counterpart of a deterministic ordinary differential equation (ODE) model \cite{Allen2012}.   Lahodny  Jr and Allen \cite{LahodnyJr2013} estimated probability of disease extinction for a Susceptible-Infected-Susceptible (SIS) multipatch model  and illustrated some differences between  thresholds for deterministic models and  stochastic models numerically. Allen and van den Driessche   \cite{Lindarelation2013} established  connections between extinction thresholds for continuous-time models and discrete-time models and illustrated the relations through numerical simulations.  Although probability of disease extinction is defined as  the probability for the number of  infections to become zero when time goes to infinity, various numerical approximations for many types of models within  finite time  showed good agreement with predicted extinction probability  using branching processes   \cite{Allen2012,Lindarelation2013, LahodnyJr2013}.

  Deriving  relationships between  reproduction numbers and extinction  thresholds is a complex task for vector-borne diseases transmitted on heterogeneous networks due to too many parameters and large size  matrices. According to current knowledge, very little research has studied  it. The objectives of this  research are to relate the  extinction threshold, $E_0$ in a stochastic setting and the reproduction number, $R_0$ in a deterministic setting  for  vector-host meta-population models, as well as gain understandings in how to increase extinction probability.  

The contribution of our work is summarized as follows.
\begin{enumerate}
\item  Relationships between extinction thresholds and the reproduction numbers are derived for network-based vector-host models under some assumptions.
\item Numerical simulations show that the relationships still exist after removing above assumptions.
\item Consistent trends of    extinction probability  varying with disease parameters are observed through extensive numerical simulations.
\item The key parameters in predicting uncertainty of the extinction threshold     are identified using Latin Hypercube Sampling/Partial Rank Correlation Coefficient (LHS/PRCC).
\item The relationship between varying disease parameters and  potential  mitigation strategies is   biologically interpreted.
\end{enumerate}

This paper is organized as follows. Section $ \ref{section:R0}$ reviews the next generation matrix approach for computing $R_0$ and  the branching process  for deriving $E_0$.
 Section $\ref{sec:SISmodel}$ calculates $R_0$ for a deterministic vector-host model in which transmission dynamics of vectors are described by a Susceptible-Infected (SI) model and transmission dynamics  of hosts are described by an SIS model.  We relate $E_0$ of  corresponding CTMC model  and   $R_0$ analytically.
      In Section $\ref{sec:SEIRmodel}$,  an analogue of results in Section \ref{sec:SISmodel} is obtained
      for a model in which transmission dynamics of vectors are  described by a  Susceptible-Exposed-Infected (SEI) model and transmission dynamics of hosts are described by a Susceptible-Exposed-Infected-Recovered (SEIR)  model.
      Local transmission and trans-location transmission due to proximity  for  vector-borne diseases  are both considered in Sections $\ref{sec:SISmodel}$ and $\ref{sec:SEIRmodel}$. In Section $\ref{sec:NumericalExample}$, the relationships derived in Sections $\ref{sec:SISmodel}$ and $\ref{sec:SEIRmodel}$ are numerically shown to hold  without any assumptions  for simplified malaria and Rift Valley fever meta-population models. The sensitivity  test sorted out the key parameters in predicting  uncertainty of extinction probability. Relationships between varying parameters and extinction probabilities are explored through extensive simulations for homogeneous populations and a two-node network.  Section $ \ref{section:result}$ provides a  summary and discussion of mathematical derivations and simulation results.

\section{Preliminary}
\label{section:R0}
The next generation matrix approach  used to compute $R_0$ for compartmental models  is reviewed here, followed by a  review of the multitype  branching process approximation  used to derive $E_0$ for corresponding CTMC models.
\subsection{Computation of $R_0$ using the next generation matrix approach}

  The  next generation matrix approach is frequently used to compute $R_0$.
  In this section, we quickly review this approach. For more details, we refer to \cite[Chapter 5]{diekmann2000mathematical}, \cite{van2002reproduction}.
     For simplicity, let $Y_i, i=1,\cdots, m$ stand for compartments that are only  related
to infected and asymptomatically infected individuals.
 The original nonlinear system of ODEs including these compartments can be written as
 $\frac{\partial Y_i}{\partial t}=\mathscr{F}-\mathscr{V}$, where $\mathscr{F}=(\mathscr{F}_i)$ and $\mathscr{V}=(\mathscr{V}_i)$ represent new infections and transfer between compartments, respectively.
Moreover,  $\mathscr{F}_i $  represents the rate at which  new infections appear in compartment $i$, and   $\mathscr{V}_i=\mathscr{V}_i^{-}-\mathscr{V}_i^{+}$, where $\mathscr{V}_i^{-}$ (resp. $\mathscr{V}_i^{+}$) represents the rate at which individuals  transfer  from (resp. into) compartment $i$.   The
Jacobian matrices $F$ representing transmission, and $V$ representing transition  are defined as:
\begin{equation}\label{Jacobian}
F= [\frac{\partial \mathscr{F}_i (x^0)}{\partial x_j}], \quad V= [\frac{\partial \mathscr{V}_i(x^0)}{\partial x_j}],
\end{equation}
where $x^0$ denotes  disease free equilibrium (DFE), and $x_j$ is the number or proportion of infected individuals in compartment $j$, where $j=1,  \cdots, m$. Matrix $F$  is nonnegative and $V$ is  a  nonsingular M-matrix.

Matrix $FV^{-1}$ is called the next generation matrix. The  $(i, k)$ entry of $FV^{-1}$ indicates the expected number of new infections  in compartment $i$ produced by the infected individual originally introduced  into compartment $k$, where $i, k=1, \cdots, m$.

The {\it reproduction number}, $R_0$, is defined as  the spectral radius of  $FV^{-1}$, denoted by $\rho(FV^{-1})$.

\subsection{Deriving $E_0$ using branching process approximation }
\label{sec:threshold}
Calculating the probability of disease extinction is one of the most interesting applications of branching process.  The branching process may lead  to disease extinction or persistence. We are interested  in the conditions under which a disease may become extinct and the probability for this event to occur. First, we review the approach of using branching process to compute extinction threshold and extinction probability for multi-type infections.

We refer to  \cite{ Allen2012,Penisson2010}  for the rest of this section.
 Let $ \overrightarrow  X(t)=(X_1(t),\cdots,$ $X_n(t))^{T}: t \in (0,  \infty) $ be a set of discrete-valued vector random variables. Assume that  individuals of type $i$ produce individuals of type $j$ and that the number of infected individuals produced by type $i$ are independent of  the number of infected individuals produced by other individuals of type $i$ or type $j$ for  $i, j=1, \cdots, n, i\neq j$. Additionally, individuals of type $i$ have identical probability generating function (pgf).  Let $\{X_{ji}\}_{j=1}^n$ be the offspring random variables for type $i$, where  $X_{ji}$ is  the number of infected individuals of type $j$ produced by individuals of type $i$. The probability that one individual of type $i$ produces $x_j$ infected individuals of type $j$  is given as  $$P_i(x_1,\cdots, x_n)={\rm Prob}\{X_{1i}=x_1, \cdots, X_{ni}=x_n \}.$$

The offspring pgf array   $(g_1,  \cdots, g_n): [0, 1]^n \rightarrow [0,1]^n$,  is defined as
\begin{align} g_i(w_1,\cdots, w_n)=\sum_{x_n=0}^\infty \cdots \sum_{x_1=0}^\infty P_i(x_1,\cdots, x_n)w_1^{x_1}\cdots w_n^{x_n}. \label{equation:gi} \end{align}
Note that a trivial  fixed point of $ (g_1,  \cdots, g_n)$ always exists at $\mathbf{1}=(1,  \cdots, 1)$.

We denote    by $M=[m_{ij}]_{n\times n}$ the expectation matrix of offspring distribution which is nonnegative, where $m_{ij}:=\frac{\partial g_i}{\partial w_j}
|_{x=1}<\infty$ represents the expected number of  new infected individuals of type $j$ produced by an individual of type $i$.

 The {\it extinction threshold}, $E_0$ is defined as the spectral radius of the expectation matrix, denoted by $\rho(M)$.

Recall that $(B_0)$ and $(B_1)$ assumptions in \cite{Penisson2010} are as follows.
\begin{enumerate}
\item[$(B_0)$]  $g_i$ is not simple. Here, a function is called simple if it is  linear with no constant term.
\item[$(B_1)$]  Matrix $M$ is irreducible.
\end{enumerate}
If $E_0>1$,  under assumptions $(B_0)$ and $(B_1)$, the pgf  has  at most one fixed point in $(0, 1)^n$, denoted by $w^*=(w_1^*, \cdots, w_n^*)$, if  extinction array $w^*$ in $(0, 1)^n$ exists.  In the following,  extinction array only refer to  $w^* \in (0, 1)^n$.  If $I_j(0)=i_j$, then disease extinction probability, denoted by $P_{E}$, is
\begin{align}
P_{E}=\lim _{t \rightarrow \infty} {\rm Prob}\{\overrightarrow I(t)=0\}=w_1^{*i_1}\cdots w_n^{*i_n}<1.  \label{extprop}
\end{align}

  If $E_0 \leq 1$,  then  $$P_{E}=\lim _{t \rightarrow \infty} {\rm Prob}\{\overrightarrow I(t)=0\}=1.$$
\section{SI vector model and SIS host metapopulation model}
\label{sec:SISmodel}
In this section, a deterministic vector-host model in which disease transmission dynamics of vectors are described by an SI model, while transmission dynamics of hosts are described by an SIS model. The reproduction number and extinction threshold for corresponding CTMC model are analytically related.


\subsection{The reproduction number}\label{SImodelR0}
The model for vectors consists of compartment $G$ representing  susceptible vectors,  and compartment $J$ representing  infected vectors.  Disease dynamics of hosts are modeled by an SIS model.
\allowdisplaybreaks
\begin{equation}
\begin{aligned}
\frac{\dif G_{i}}{\dif t} &=\eta_{i}-\beta_{i}G_{i}I_{i}/N_{i}-\sum^n_{j=1, j \neq
i}\omega_{ji}G_{i}I_{j}/N_{j}-\mu_{i}G_{i} \\
\frac{\dif J_{i}}{\dif t} &=\beta_{i}G_{i}I_{i}/N_{i}+\sum^n_{j=1, j \neq
i}\omega_{ji}G_{i}I_{j}/N_{j}-\mu_{i}J_{i}\\
\frac{\dif S_{i}}{\dif t} &=\psi_{i}+\gamma_{i}I_{i}-\alpha_{i}S_{i}J_{i}/N_{i} -\sum^n_{j=1, j \neq
i}\sigma_{ji}S_{i}J_{j}/N_{i}-d_{i}S_{i}\\
\frac{\dif I_{i}}{\dif t} &=\alpha_{i}S_{i}J_{i}/N_{i}+\sum^n_{j=1, j \neq
i}\sigma_{ji}S_{i}J_{j}/N_{i}
-\gamma_{i}I_{i}-d_{i}I_{i}
 \end{aligned}
 \label{equation:SImodel}
\end{equation}
The recruitment rate  of  vectors (resp. hosts) in node $i$  is $\eta_{i}$ (resp. $\psi_i$) for all  $i=1, \cdots, n$. The rate of new infections in vectors in node $i$  produced by local hosts, and hosts in other nodes are  $\beta_{i}G_{i}I_{i}/N_{i}$ and   $\sum^n_{j=1, j \neq
i}\omega_{ji}G_{i}I_{j}/N_{j}$, respectively. The death rate of susceptible and infected vectors in node $i$ are $\mu G_{i}$ and $\mu J_{i}$, respectively.
 The rate of host infection   in node $i$  produced by local vectors, and vectors in other nodes are  $\alpha_{i}S_{i}J_{i}/N_{i} $ and  $\sum^n_{j=1, j \neq
i}\sigma_{ji}S_{i}J_{j}/N_{i}$, respectively. The death rates of susceptible and infected hosts  in node $i$  are $d_iS_i$ and $d_i I_i$, respectively. The rate of  recovery  for hosts in node $i$  is $\gamma_{i}I_{i} $.

Since $J_i$ and $I_i$, $i=1, \cdots n$ are only  compartments related
to infected and asymptomatically infected,  system of ODEs   $(\ref{equation:SImodel})$ can be rewritten as follows.
\begin{equation*}
\frac{d}{dt}\left[
\begin{array}{rllllllllllllllllllllllllllllllllllllllll}
J_{1}  &\cdots & J_{n}& I_{1} &\cdots & I_{n}
\end{array}\right]^T
= \mathscr{F}- \mathscr{V}.
\end{equation*}

A unique solution at DFE, represented by  $(G_i^0, 0, N_i^0, 0)$ exists, where $G_{i}^0=\frac{\eta_{i}}{\mu_{i}}$ and $N_{i}^0=\frac{\psi_{i}}{d_{i}}$. The Jacobian matrices $F$ and $V$ defined in (\ref{Jacobian}) for this model are
$$F=\begin{bmatrix}
  0 &\mathcal A \\
 \mathcal B &0
\end{bmatrix},\quad
V=\begin{bmatrix}
  \Lambda_1& 0\\
  0 &\Lambda_2
\end{bmatrix},$$
where
\begin{equation}\label{eqAB}
\mathcal A=\begin{bmatrix}
\hat\beta_{1}&\hat\omega_{21}&\cdots&\hat\omega_{n1}\\
\hat\omega_{12}&\hat\beta_{2}&\cdots&\hat\omega_{n2}\\
\cdots  &\cdots &\ddots &\cdots\\
\hat\omega_{1n}&\hat\omega_{2n}&\cdots&\hat\beta_{n}\\
\end{bmatrix},
\quad
\mathcal B=\begin{bmatrix}
\alpha_{1}&\sigma_{21}&\cdots&\sigma_{n1}\\
\sigma_{12}&\alpha_{2}&\cdots&\sigma_{n2}\\
\cdots  &\cdots &\ddots &\cdots\\
\sigma_{1n}&\alpha_{2}&\cdots&\alpha_{n}
\end{bmatrix},
\end{equation}

\begin{equation}\label{Lambda1and2}
\begin{split}
  \Lambda_1&={\rm diag}(\mu_1, \cdots, \mu_n),\quad
  \Lambda_2={\rm diag}(d_1+\gamma_1,\cdots, d_n+\gamma_n).
\end{split}
\end{equation}
Here 
$$\textstyle \hat{\beta}_i=\frac{\beta_iG_i^0}{N_i^0}\quad {\rm and}\quad \hat{\omega}_{ij}=\frac{\omega_{ij}G_j^0}{N_i^0}.$$
 The notation ${\rm diag}(\mu_1,\mu_2,\cdots, \mu_n)$ represents the diagonal matrix with diagonal entries $\mu_1, \cdots, \mu_n$.
To calculate  $R_0$, we first prove the following lemma.
\begin{lemma}\label{lem1}
Let $A_1, A_2$ be square matrices of the same size and $A=\begin{bmatrix}
0&A_1\\
A_2 & 0
\end{bmatrix}$, then
$\rho(A)
=\sqrt{\rho(A_2A_1)}.$
\end{lemma}

\begin{proof}
For any $\lambda \neq 0$,
\begin{equation}\label{eq1}
\begin{split}
|\lambda I-A|= &\left| {%
\begin{array}{cccccccccccc}
\lambda I&-A_1\\
-A_2& \lambda I \\
\end{array}
}\right|
=\left|
\begin{array}{cccccccccccc}
 \lambda I &-A_1\\
0& \lambda I-\frac{A_2A_1}{\lambda} \\
\end{array}
\right|=|\lambda^2I-A_2A_1|.
\end{split}
\end{equation}
Therefore, $\rho(A)=\sqrt{\rho(A_2A_1)}$ if $\rho(A_2A_1) \neq 0$.

If $\rho(A_2A_1)= 0$, we assume  that $\rho(A) \neq 0$. Then there exists a $\lambda' \neq 0$ such that
$|\lambda'I-A|=0$.
By (\ref{eq1}),
$|\lambda'^2I-A_2A_1|=0$ for a nonzero $\lambda'$, contradicting the assumption that $\rho(A_2A_1)=0$.
Therefore, $\rho(A)=\sqrt{\rho(A_2A_1)}.$ 
\end{proof}
A direct calculation gives
$FV^{-1}=\begin{bmatrix}
  0 &\mathcal A\Lambda_2^{-1}\\
   \mathcal B\Lambda_1^{-1} &0\end{bmatrix}.$
By Lemma \ref{lem1}, we have the following proposition.
\begin{proposition}
\label{prop1}
The reproduction number of the model (\ref{equation:SImodel}) is
\begin{align}
R_0=\sqrt{\rho( \mathcal B\Lambda_1^{-1}\mathcal A \Lambda_2^{-1} )}.  \label{equation:deterministicR0}
\end{align}
\end{proposition}
\subsection{The threshold for extinction probability}
In this section, we compute $E_0$  for  corresponding  CTMC  of  model $(\ref{equation:SImodel})$. See Table $\ref{table:vectorhosttransitions}$ for state transitions and rates.

The pgfs are:
\begin{equation*}
g_i(w_1,\cdots, w_n, u_1,\cdots, u_n)= \left\{ \begin{array}{ll}
\frac{\alpha_{i}w_iu_i+\sum_{j=1, j \neq i}^n\sigma_{ij}w_iu_j+\mu_{i}}{\alpha_{i}+\sum_{j=1, j \neq i}^n\sigma_{ij}+\mu_{i}},
&{\rm if}\   1 \leq i \leq n,\\
\frac{\hat\beta_{k}u_kw_k+\sum_{j=1, j \neq k}^n\hat\omega_{kj}u_kw_j+d_{k}+\gamma_k}{\hat\beta_{k}+\sum_{j=1, j \neq k}^n\hat\omega_{kj}+d_{k}+\gamma_k},
 &{\rm if}\  n+1 \leq i \leq 2n,
\end{array}\right.
\end{equation*}
where $j=1, \cdots, n$, the index $k=i-n$ for  $n+1 \leq i \leq 2n$,
$w_i$ represents $I_{V_i}=1,  I_{H_i}=0$,
and $u_i$ represents $I_{H_i}=1,  I_{V_i}=0 \ \ {\rm for}\ i=1, \cdots, n$.

\begin{table}
\centering
\begin{tabular}{|c|c|c|}
\hline
Description &State transition $ a \rightarrow b $ & Rate $P(a, b)$\\
\hline
Host birth & $(S, I, G, J) \rightarrow (S+1, I, G, J)$ & $\psi$\\
\hline
Death of $S$ & $(S, I, G, J) \rightarrow (S-1, I, G, J)$ & $dS$\\
\hline
Host local infection & $(S, I, G, J) \rightarrow (S-1, I+1, G, J)$ & $\alpha SJ/N$\\
\hline
Host infection by $J_j$& $(S, I, G, J) \rightarrow (S-1, I+1, G, J)$ & $\sigma_{ji} S_iJ_j/N_i$\\
\hline
Host recovery & $(S, I, G, J) \rightarrow (S+1, I-1, G, J)$ & $\gamma I$\\
\hline
Death of $I$ & $(S, I, G, J) \rightarrow (S, I-1, G, J)$ & $dI$\\
\hline
Vector birth & $(S, I, G, J) \rightarrow (S, I, G+1, J)$ & $\eta$\\
\hline
Death of $G$ & $(S, I, G, J) \rightarrow (S, I, G-1, J)$ & $\mu G$\\
\hline
Vector local infection & $(S, I, G, J) \rightarrow (S, I, G-1, J+1)$ & $\beta GI/N$\\
\hline
Vector  infection by $I_j$& $(S, I, G, J) \rightarrow (S, I, G-1, J+1)$ & $\omega_{ji} G_iI_j/N_j$\\
\hline
Death of $J$ & $(S, I, G, J) \rightarrow (S, I, G, J-1)$ & $\mu J$\\
\hline
\end{tabular}
\caption{State transitions and rates for corresponding  continuous-time Markov chain for deterministic model  $(\ref{equation:SImodel})$  omitting node index $i$.}
\label{table:vectorhosttransitions}
\end{table}

The expectation matrix $M$ is:
\begin{equation}\label{eqM}
M= \left[ {%
\begin{array}{cccccccccccc}
\Lambda_3\Lambda_4&\mathcal A \Lambda_5\\
\mathcal B \Lambda_4&\Lambda_6\Lambda_5\\
\end{array}
}\right],
\end{equation}
where $\mathcal A, \mathcal B$ are the same as those in (\ref{eqAB}), and
\begin{equation*}
\begin{split}
  \Lambda_3&={\rm diag}(\alpha_1+\sum_{i\neq 1} \sigma_{1i},\cdots,\alpha_n+\sum_{i\neq n}\sigma_{ni}), \quad
  \Lambda_4={\rm diag}(\frac{1}{C_1}, \cdots, \frac{1}{C_n}),\\
\Lambda_6&={\rm diag}(\hat{\beta}_1+\sum_{i\neq 1}\hat{\omega}_{1i},\cdots, \hat{\beta}_n+\sum_{i\neq n}\hat{\omega}_{ni}), \quad
\Lambda_5={\rm diag}(\frac{1}{D_1}, \cdots, \frac{1}{D_n}),\\
C_i&=\alpha_{i}+\sum_{j\neq i}\sigma_{ij}+\mu_{i},\quad D_i=\hat{\beta_{i}}+\sum_{j\neq i}\hat{\omega}_{ij}+d_{i}+\gamma_{i},\quad
{\rm for}\ i=1, \cdots,n.
\end{split}
\end{equation*}
Note that if both $\mathcal A$ and $\mathcal B$ are positive matrices, then the assumptions $(B_0)$ and $(B_1)$  in \cite{Penisson2010} hold for this model.
\begin{lemma}\label{lem2}
  Let $A_1, A_2$ be nonnegative square matrices with the same size such that $\rho(A_2A_1)$ is an eigenvalue of $A_2A_1$ and
    $\Lambda, \Lambda'$ be nonnegative diagonal matrices such that $0\leq k_1I \leq \begin{bmatrix} \Lambda &0 \\0&\Lambda'\end{bmatrix}\leq k_2I$ for some real numbers $k_1, k_2$. Then
  the spectral radius of
  $B=\begin{bmatrix}
    \Lambda & A_1\\
    A_2 &\Lambda'
  \end{bmatrix}$
  satisfies that
  \[ \sqrt{\rho(A_2A_1)}+k_1\leq \rho(B) \leq \sqrt{\rho(A_2A_1)}+k_2. \]
\end{lemma}
\begin{proof}
Since
\(0\leq \begin{bmatrix}
   k_1I& A_1\\
  A_2 & k_1I
\end{bmatrix} \leq B \leq \begin{bmatrix}
   k_2I& A_1\\
  A_2 & k_2I
\end{bmatrix},\)
 by Theorem $4$ in \cite{LingXue2013MB},
 \begin{equation}\label{eqrhoM}
 \rho(\begin{bmatrix}
   k_1I& A_1\\
  A_2 & k_1I
\end{bmatrix}) \leq \rho(B) \leq \rho(\begin{bmatrix}
   k_2I& A_1\\
  A_2 & k_2I
\end{bmatrix}).
 \end{equation}
  By hypothesis and (\ref{eq1}), $\rho(\begin{bmatrix}
   0&A_1\\
   A_2&0
  \end{bmatrix})$ is an eigenvalue of $\begin{bmatrix}
   0&A_1\\
   A_2&0
  \end{bmatrix}$.
Following the fact that $|\lambda'+k|<\lambda +k$ for any $k>0$ if $|\lambda'|<\lambda$,
\[\rho(\begin{bmatrix}
   k_1I& A_1\\
  A_2 & k_1I
\end{bmatrix})=\rho(\begin{bmatrix}
0& A_1\\
  A_2 & 0
\end{bmatrix})+k_1=\sqrt{\rho(A_2A_1)}+k_1.\]
Similarly, $\rho(\begin{bmatrix}
   k_2I& A_1\\
  A_2 & k_2I
\end{bmatrix})=\sqrt{\rho(A_2A_1)}+k_2.$
Lemma follows (\ref{eqrhoM}) and Lemma \ref{lem1}.
\end{proof}
\begin{remark}\label{remark1}
If both $A_1$ and $A_2$ are positive matrices, then $\rho(A_2A_1)$ is an eigenvalue of $A_2A_1$ by Perron-Frobenius theorem.
\end{remark}
By Lemma \ref{lem2}, we have the following proposition.
\begin{proposition}
\label{prop2}
 The extinction threshold of model (\ref{equation:SImodel}) satisfies that
\begin{equation}
  \begin{split}
& \min_{1\leq i\leq n}(\frac{\alpha_{i}+\sum_{j=1, j \neq i}^n\sigma_{ij}}{C_i}, \frac{\hat\beta_{i}+\sum_{j=1, j \neq i}^n\hat\omega_{ij}}{D_i}  )+\sqrt{\rho(\mathcal B \Lambda_5\mathcal A \Lambda_4)} \leq E_0 \nonumber \\
&\leq \max_{1\leq i\leq n}(\frac{\alpha_{i}+\sum_{j=1, j \neq i}^n\sigma_{ij}}{C_i}, \frac{\hat\beta_{i}+\sum_{j=1, j \neq i}^n\hat\omega_{ij}}{D_i}  )+\sqrt{\rho(\mathcal B \Lambda_5\mathcal A \Lambda_4)}\label{equation:rhoWbounds}.
  \end{split}
\end{equation}
\end{proposition}
\subsection{The relationship between $R_0$ and $E_0$}
\label{sec:relationship}
To obtain a theoretical relationship between  $R_0$ in $(\ref{equation:deterministicR0})$ and $E_0$, we assume that
\begin{equation}\label{eq-assumption}
  \frac{\mu_i}{C_i}=k_1\quad {\rm and}\quad \frac{d_i+\gamma_i}{D_i}=k_2,\quad \forall \ i=1,  \cdots, n
\end{equation}
for constant numbers $k_1, k_2\in [0,1]$ throughout this section. The assumption  can be interpreted biologically as: the probability of natural death is identical  for vectors from each node, and the probability of natural death is identical for hosts from each node.
The assumption shall be removed for numerical simulations in the next section.
\begin{theorem}\label{thm1}
  Under the assumption (\ref{eq-assumption}),
  \begin{itemize}
    \item[(1)] If $R_0 \leq \frac{1-k_2}{1-\sqrt{k_1k_2}}\leq 1$ or $E_0 \leq \frac{1-k_2}{1-\sqrt{k_1k_2}}\leq 1$, then $R_0 \leq E_0$;

 \item[(2)] If $R_0 \geq \frac{1-k_1}{1-\sqrt{k_1k_2}}\geq 1$ or $E_0 \geq \frac{1-k_1}{1-\sqrt{k_1k_2}}\geq 1$, then $R_0 \geq E_0$.


  \end{itemize}
\end{theorem}
\begin{proof}
Under the assumption (\ref{eq-assumption}),
$\Lambda_1\Lambda_4=k_1I$, $\Lambda_3\Lambda_4=(1-k_1)I$, $\Lambda_2\Lambda_5=k_2I$ and $\Lambda_6\Lambda_5=(1-k_2)I$,
where $I$ is the identity matrix.
 Therefore, $M$ in (\ref{eqM}) can be rewritten as follows,
\begin{eqnarray*}
  M
  =&\begin{bmatrix}
    0 & k_2 \mathcal A \Lambda_2^{-1}\\
    k_1 \mathcal B \Lambda_1^{-1} & 0
  \end{bmatrix}+
  \begin{bmatrix}
    (1-k_1)I & 0\\
    0 & (1-k_2)I
  \end{bmatrix}.
\end{eqnarray*}

Without loss of generality, we assume that $k_1 <k_2$.
By Lemma \ref{lem2}
and (\ref{equation:deterministicR0}),
\begin{eqnarray}\label{eq3}
R_0\sqrt{k_1k_2}+1-k_2 \leq E_0 \leq R_0\sqrt{k_1k_2}+1-k_1.
\end{eqnarray}
Following (\ref{eq3}),
\begin{equation*}\label{eq7}
\begin{split}
  R_0(1-\sqrt{k_1k_2})-(1-k_1) &\leq R_0-E_0 \leq R_0(1-\sqrt{k_1k_2})-(1-k_2),\quad \\
  \frac{1}{\sqrt{k_1k_2}}(E_0(1-\sqrt{k_1k_2})-(1-k_1)) &\leq R_0-E_0 \leq \frac{1}{\sqrt{k_1k_2}}(E_0(1-\sqrt{k_1k_2})-(1-k_2)).
\end{split}
\end{equation*}
Theorem follows the above two inequalities.
%
\end{proof}

\begin{corollary}\label{cor1}
  If the further  assumption is made that $k_1=k_2$ except  assumption (\ref{eq-assumption}), then
%
%
 $R_0\leq 1$ if and only if $E_0\leq 1$. Moreover, $|R_0-1|\geq |E_0-1|$.
\end{corollary}
\begin{proof}
  By Theorem \ref{thm1} (1), if $R_0\leq 1$, then $R_0\leq E_0$.
  Assuming that $E_0>1$, by Theorem \ref{thm1} (2), $ R_0\geq E_0$, which  is a contradiction.
  Conversely, if $E_0>1$, then $R_0\leq 1$ following a similar argument.
Hence,   $R_0\leq 1$ if and only if $E_0\leq 1$.
This proves the first part.
The second part  directly follows Theorem \ref{thm1}.
\end{proof}

\section{SEI vector model and SEIR host metapopulation model}
\label{sec:SEIRmodel}
A deterministic model  in which vectors are divided into compartments $S, E$, and $I$, and hosts are classified into compartments $S, E, I$, and $R$ is presented.  The reproduction number for this model and the extinction threshold for corresponding CTMC model are connected.
\subsection{The reproduction number}
The following model extends the model in Section $\ref{SImodelR0}$ by adding compartment $Z$ for exposed vectors, and compartment $E$ for exposed hosts. Other terms have identical  meanings as corresponding ones in model $(\ref{equation:SImodel})$. The rate at which  exposed vectors and exposed hosts in node $i$ transfer to  infected compartments are $\varphi_{i}Z_{i}$ and $\varepsilon_{i}E_{i}$, respectively.
\allowdisplaybreaks
\begin{equation}
\begin{aligned}
\frac{\dif G_{i}}{\dif t} &=\eta_{i}-\beta_{i}G_{i}I_{i}/N_{i}-\sum^n_{j=1, j \neq
i}\omega_{ji}G_{i}I_{j}/N_{j}-\mu_{i}G_{i} \\
\frac{\dif Z_{i}}{\dif t} &=\beta_{i}G_{i}I_{i}/N_{i}+\sum^n_{j=1, j \neq
i}\omega_{ji}G_{i}I_{j}/N_{j}-\varphi_{i}Z_{i}-\mu_{i}Z_{i}\\
\frac{\dif J_{i}}{\dif t} &=\varphi_{i}Z_{i}-\mu_{i}J_{i}\\
\frac{\dif S_{i}}{\dif t} &=\psi_{i}-\alpha_{i}S_{i}J_{i}/N_{i} -\sum^n_{j=1, j \neq
i}\sigma_{ji}S_{i}J_{j}/N_{i}-d_{i}S_{i}\\
\frac{\dif E_{i}}{\dif t} &=\alpha_{i}S_{i}J_{i}/N_{i}+\sum^n_{j=1, j \neq
i}\sigma_{ji}S_{i}J_{j}/N_{i}-\varepsilon_{i}E_{i}
-d_{i}E_{i}\\
\frac{\dif I_{i}}{\dif t} &=\varepsilon_{i}E_{i}-\gamma_{i}I_{i}
-d_{i}I_{i}\\
\frac{\dif R_{i}}{\dif t} &=
\gamma_{i}I_{i}-d_{i}R_{i}
\end{aligned}
\label{SEIRmodel}
\end{equation}

Compartments related
to infected and asymptomatically infected are $Z_i, E_i, J_i$ and $I_i$, $i=1,\cdots, n$. The unique solution at  DFE is $(G_i^0, 0, 0, N_i^0, 0, 0, 0)$, where 
$G_i^0$ and $N_i^0$ are the same as those in Section \ref{SImodelR0}.
The above system of ODEs  including these compartments can be rewritten as follows.
\begin{equation*}
\frac{d}{dt}\left[
\begin{array}{rllllllllllllllllllllllllllllllllllllllll}
Z_{1} &\cdots &Z_{n}  &  E_{1} \cdots &  E_{n}&  J_{1} &\cdots &  J_{n}& I_{1} &\cdots & I_{n}
\end{array}\right]^T
= \mathscr{F}- \mathscr{V}.
\end{equation*}

The Jacobian matrices $F$ and $V$ at DFE are
\begin{equation*}
F=\begin{bmatrix}
0&0&0&\mathcal{A}\\
0&0&\mathcal{B}&0\\
0&0&0&0\\
0&0&0&0\\
\end{bmatrix}, \ \ \
V=\begin{bmatrix}
\Lambda_7 &0&0&0\\
0&\Lambda_8&0&0\\
-\Lambda_9 &0&\Lambda_{1} &0\\
0&-\Lambda_{10}&0&\Lambda_{2} \\
\end{bmatrix},
\end{equation*}
where $\Lambda_{1}$ and $\Lambda_{2}$ are given in $(\ref{Lambda1and2})$;
matrices $\mathcal{A}$ and  $\mathcal{B}$ are in Equation $(\ref{eqAB})$; and
\begin{equation*}
\begin{split}
 \Lambda_7&={\rm diag}(\varphi_{1}+\mu_{1},\cdots, \varphi_{n}+\mu_{n}), \quad
 \Lambda_8={\rm diag}(\varepsilon_{1}+d_{1}, \cdots, \varepsilon_{n}+d_{n}),\\
  \Lambda_9&={\rm diag}(\varphi_{1}, \cdots, \varphi_{n} ), \quad
  \Lambda_{10}={\rm diag}(\varepsilon_{1}, \cdots, \varepsilon_{n}).\\
\end{split}
\end{equation*}
By a direct calculation,
\begin{equation*}
FV^{-1}=\begin{bmatrix}
0 &\mathcal A \Lambda_{2}^{-1} \Lambda_{10}\Lambda_8^{-1}&0&\mathcal A\Lambda_2^{-1}\\
\mathcal B\Lambda_{1}^{-1} \Lambda_9 \Lambda_7^{-1}&0 &\mathcal B \Lambda_1^{-1}&0\\
0&0&0  &0\\
0&0&0&0 \\
\end{bmatrix}.
\end{equation*}
Following Lemma $\ref{lem1}$,
\begin{proposition}
The reproduction number of the model (\ref{SEIRmodel}) is
\begin{equation}
R_0=\sqrt{\rho(\mathcal{B}\Lambda_{1}^{-1} \Lambda_{9}\Lambda_7^{-1}\mathcal{A}\Lambda_{2}^{-1} \Lambda_{10}\Lambda_8^{-1})} \label{equation:R0SEIR}.
\end{equation}
\end{proposition}
\subsection{The threshold for  extinction probability}
State transitions and rates for corresponding CTMC  of  model  $(\ref{SEIRmodel})$  are listed  in Table \ref{table:SEIRtransitions}.
\begin{table}
\centering
\begin{tabular}{|p{105pt}|p{230pt}|p{50pt}|}
\hline
Description &State transition $ a \rightarrow b $ & Rate $P(a, b)$\\
\hline
Host birth & $(S, E, I, R, G, Z, J) \rightarrow (S+1, E, I, R, G, Z, J)$ & $\psi$\\
\hline
Death of $S$ & $(S, E, I, R, G, Z, J) \rightarrow (S-1, E, I, R, G, Z, J)$ & $dS$\\
\hline
Death of $E$ & $(S, E, I, R, G, Z, J) \rightarrow (S, E-1, I, R, G, Z, J)$ & $dE$\\
\hline
Death of $I$ & $(S, E, I, R, G, Z, J) \rightarrow (S, E, I-1, R, G, Z, J)$ & $dI$\\
\hline
Death of $R$ & $(S, E, I, R, G, Z, J) \rightarrow (S, E, I, R-1, G, Z, J)$ & $dR$\\
\hline
Host local infection & $(S, E, I, R, G, Z, J) \rightarrow (S-1, E+1, I, R, G, Z, J)$ & $\alpha SJ/N$\\
\hline
Host  infection by $J_j$ & $(S, E, I, R, G, Z, J) \rightarrow (S-1, E+1, I, R, G, Z, J)$ & $\sigma_{ji} SJ_j/N$\\
\hline
Host recovery & $(S, E, I, R, G, Z, J) \rightarrow (S, E, I-1, R+1, G, Z, J)$ & $\gamma I$\\
\hline
Host Latent to infectious  & $(S, E, I, R, G, Z, J) \rightarrow  (S, E-1, I+1, R, G, Z, J)$ & $\varepsilon E$\\
\hline
Vector birth & $(S, E, I, R, G, Z, J) \rightarrow (S, E, I, R, G+1, Z, J)$ & $\eta$\\
\hline
Death of $G$ & $(S, E, I, R, G, Z, J) \rightarrow (S, E, I, R, G-1, Z, J)$ & $\mu G$\\
\hline
Death of $Z$ & $(S, E, I, R, G, Z, J) \rightarrow S, E, I, R, G, Z-1, J)$ & $\mu Z$\\
\hline
Death of $J$ & $(S, E, I, R, G, Z, J) \rightarrow (S, E, I, R, G, Z, J-1)$ & $\mu J$\\
\hline
Vector local infection & $(S, E, I, R, G, Z, J) \rightarrow (S, E, I, R, G-1, Z+1, J)$ & $\beta GI/N$\\
\hline
Vector  infection by $I_j$& $(S, E, I, R, G, Z, J) \rightarrow  (S, E, I, R, G-1, Z+1, J)$ & $\omega_{ji} GI_j/N_j$\\
\hline
Vector Latent to infectious  & $(S, E, I, R, G, Z, J) \rightarrow  (S, E, I, R, G, Z-1, J+1)$ & $\varphi Z$\\
\hline
\end{tabular}
\caption{State transitions and rates for corresponding continuous-time Markov chain for deterministic  model  $(\ref{SEIRmodel})$  omitting node index $i$.}
\label{table:SEIRtransitions}
\end{table}
The pgfs are:
\begin{equation*}
g_i(w_1,\cdots, w_{2n}, u_1,\cdots, u_{2n})= \left\{ \begin{array}{llll}
\frac{\varphi_i  u_i+\mu_{i}}{\varphi_{i}+\mu_{i}},
&{\rm if}\   1 \leq i \leq n,\\
\frac{\varepsilon_{k}u_i+d_{k}}{\varepsilon_{k}+d_{k}},
 &{\rm if}\  n+1 \leq i \leq 2n,\\
\frac{\alpha_{p}u_pw_{p+n}+\sum_{j=1, \neq p}^{n}\sigma_{pj}u_pw_{j+n}+\mu_{p}}{\alpha_{p}+\sum_{j=1, \neq p}^{n}\sigma_{pj}+\mu_{p}},
&{\rm if}\   2n+1 \leq i  \leq 3n,\\
\frac{\hat\beta_{q}u_{q+n}w_q+\sum_{j=1, j \neq q}^n\hat\omega_{qj}u_{q+n}w_j+d_{q}+\gamma_{q}}{\hat\beta_{q}+\sum_{j=1, j \neq q}^n\hat\omega_{qj}+d_{q}+\gamma_{q}},
 &{\rm if}\  3n+1 \leq i \leq 4n,
\end{array}\right.
\end{equation*}
where $w_i$ represents only $Z_{i}=1 $, $w_{i+n}$ represents $E_{i}=1$, $u_i$ represents $J_{i}=1 $, and $u_{i+n}$ represents $I_{i}=1$ for  $i=1, \cdots, n$.
The indexes $k=i-n$ for $1 \leq i \leq n$, $p=i-2n$  for $n+1 \leq i \leq 2n$, and $q=i-3n$ for  $3n+1 \leq i \leq 4n$.

The expectation matrix $M$ is:
\begin{align*}
M=\left[  {%
\begin{array}
[c]{cccccccc}%
0&0&0&\mathcal{A}\Lambda_{5}\\
0&0&\mathcal{B}\Lambda_{4}&0\\
\Lambda_{9}\Lambda_{7}^{-1}&0&I-\Lambda_{1}\Lambda_{4}&0\\
0&\Lambda_{10}\Lambda_{8}^{-1}&0&I-\Lambda_{2}\Lambda_{5}\\
\end{array}
}\right].
\end{align*}
Similarly, the assumptions $(B_0)$ and $(B_1)$  in \cite{Penisson2010} hold for this model if both $\mathcal{A} $ and $\mathcal{B}$ are positive matrices.
By Lemmas \ref{lem1} and \ref{lem2}, as well as  Remark \ref{remark1}, we have the following proposition.
\begin{proposition}
The extinction threshold of the model (\ref{SEIRmodel}) satisfies that
\begin{eqnarray*}
&\sqrt[4]{\rho(\Lambda_{10}\Lambda_8^{-1}\mathcal B\Lambda_4\Lambda_9\Lambda_7^{-1}\mathcal A \Lambda_5)}+\min_{1\leq i\leq n}(\frac{\alpha_{i}+\sum_{j \neq i}\sigma_{ij}}{C_i}, \frac{\hat\beta_{i}+\sum_{j \neq i}\hat\omega_{ij}}{D_i}  ) \leq E_0\nonumber\\
& \leq\sqrt[4]{\rho(\Lambda_{10}\Lambda_8^{-1}\mathcal B\Lambda_4\Lambda_9\Lambda_7^{-1}\mathcal A \Lambda_5)}+\max_{1\leq i\leq n}(\frac{\alpha_{i}+\sum_{j \neq i}\sigma_{ij}}{C_i}, \frac{\hat\beta_{i}+\sum_{j \neq i}\hat\omega_{ij}}{D_i}  ).
\end{eqnarray*}
\end{proposition}
%
%
%
%
\subsection{The relationship between $R_0$ and $E_0$}

In this section, the assumption $(\ref{eq-assumption})$ holds and $k_1 <k_2$.
Under the assumption (\ref{eq-assumption}), by Lemma \ref{lem2},
\begin{equation}\label{rhoMinequality}
\begin{split}
&\sqrt[4]{k_1k_2\rho(\Lambda_{10}\Lambda_8^{-1}\mathcal B\Lambda_1^{-1}\Lambda_9\Lambda_7^{-1}\mathcal A \Lambda_2^{-1})}+1-k_2 \leq E_0\\
 &\leq \sqrt[4]{k_1k_2\rho(\Lambda_{10}\Lambda_8^{-1}\mathcal B\Lambda_1^{-1}\Lambda_9\Lambda_7^{-1}\mathcal A \Lambda_2^{-1})}+1-k_1.
\end{split}
\end{equation}
Recall that, for any square matrices $A,B$ with the same size,  $\rho(AB)=\rho(BA)$. By this property,
\begin{equation}\label{eq6}
\rho(\Lambda_{10}\Lambda_8^{-1}\mathcal B\Lambda_4\Lambda_9\Lambda_7^{-1}\mathcal A \Lambda_2^{-1})=\rho(\mathcal B\Lambda_4\Lambda_9\Lambda_7^{-1}\mathcal A \Lambda_2^{-1}\Lambda_{10}\Lambda_8^{-1}).
\end{equation}
By $(\ref{equation:R0SEIR})$, $(\ref{rhoMinequality})$ and (\ref{eq6}),
\begin{eqnarray*}\label{eq4}
\sqrt{R_0}\sqrt[4]{k_1k_2}+1-k_2 \leq E_0 \leq \sqrt{R_0}\sqrt[4]{k_1k_2}+1-k_1.
\end{eqnarray*}
Hence,
\begin{equation*}
\begin{split}
 \sqrt{R_0}(1-\sqrt[4]{k_1k_2})-(1-k_1) &\leq \sqrt{R_0}-E_0 \leq \sqrt{R_0}(1-\sqrt[4]{k_1k_2})-(1-k_2),\\
 \frac{1}{\sqrt[4]{k_1k_2}}(E_0(1-\sqrt[4]{k_1k_2})-(1-k_1)) &\leq \sqrt{R_0}-E_0 \leq \frac{1}{\sqrt[4]{k_1k_2}}(E_0(1-\sqrt[4]{k_1k_2})-(1-k_2)).
\end{split}
\end{equation*}

Similarly,  we have the following theorem.
\begin{theorem}\label{thm2}
  Under assumption $(\ref{eq-assumption})$,
  \begin{itemize}
    \item[(1)] If $\sqrt{R_0} \leq \frac{1-k_2}{1-\sqrt[4]{k_1k_2}}\leq 1$ or $E_0 \leq \frac{1-k_2}{1-\sqrt[4]{k_1k_2}}\leq 1$,
     then $\sqrt{R_0} \leq E_0$;

 \item[(2)] If $\sqrt{R_0} \geq \frac{1-k_1}{1-\sqrt[4]{k_1k_2}}\geq 1$ or $E_0 \geq \frac{1-k_1}{1-\sqrt[4]{k_1k_2}}\geq 1$,
  then $\sqrt{R_0} \geq E_0$.


  \end{itemize}
\end{theorem}
\begin{corollary}\label{cor2}
  If a further assumption is made that $k_1=k_2$ besides  assumption (\ref{eq-assumption}), then
%
%
$\sqrt{R_0} \leq 1$ if and only if $E_0\leq 1$. Furthermore, $|\sqrt{R_0}-1|\geq |E_0-1|$.
\end{corollary}
\begin{proof}
The proof is similar to that of Corollary \ref{cor1}.
%
%
\end{proof}
\section{Numerical results}
We show  numerically the general relations between $R_0$ and $E_0$ for two models on heterogeneous networks.  Significant  parameters in predicting the uncertainly in $E_0$ are found by  Latin Hypercube Sampling/Partial Rank Correlation Coefficient (LHS/PRCC) analysis. Finally, the trends of parameters varying with extinction array is summarized.

\label{sec:NumericalExample}
\subsection{Numerical results on relations between $R_0$ and $E_0$}
\begin{figure}[!htbp]
\centering
\subfigure[When $R_0\leq 1 $,  $R_0 \leq  E_0$.]{
\label{fig:SIR0<1}
\includegraphics[angle=0,width=6.1cm,height=5.9cm]{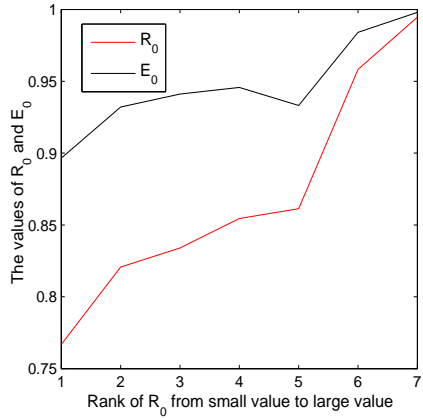}}
\hspace{0.1in}
\subfigure[When $R_0\geq 1 $,  $R_0 \geq  E_0$.]{
\label{fig:SIR0>1}
\includegraphics[angle=0,width=6.1cm,height=5.9cm]{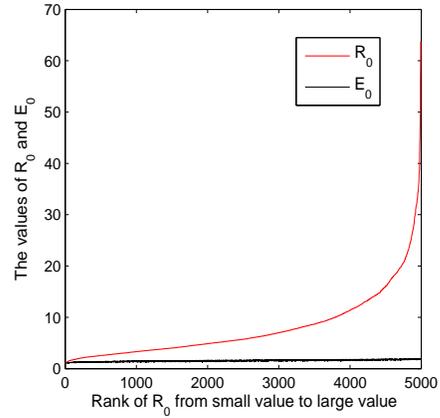}}
\hspace{0.1in}
\subfigure[When $E_0 \leq 1 $,  $R_0 \leq  E_0$.]{
\label{fig:SIrhoM<1}
\includegraphics[angle=0,width=6.1cm,height=5.9cm]{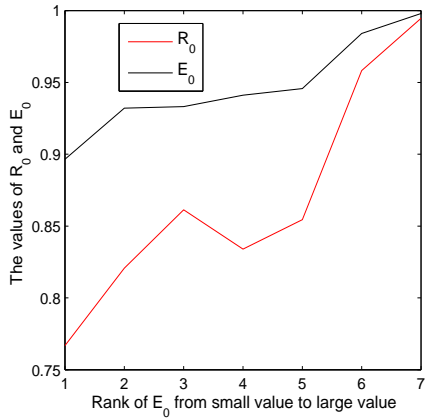}}
\hspace{0.1in}
\subfigure[When $E_0 \geq 1 $,  $R_0 \geq  E_0$.]{
\label{fig:SIrhoM>1}
\includegraphics[angle=0,width=6.1cm,height=5.9cm]{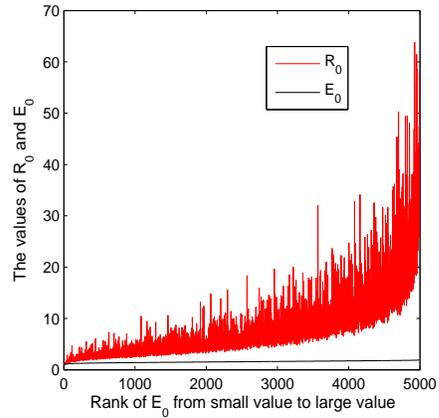}}
\hspace{0.1in}
\caption{Relationships between $R_0$ and $E_0$ for malaria model.}
\label{fig:malariasimulation}
\end{figure}

Model $(\ref{equation:SImodel})$  is applied  to study thresholds for malaria transmission through numerical simulations. Five thousand realizations   with parameters uniformly distributed within the ranges listed  in Table $\ref{table:malariaparameters}$ on a  four-node network give rise to  $R_0$ ranging from $0.7668$ to  $63.8111$ and $E_0$ from $0.8965$ to $1.9140$. The ranges of $R_0$ and $E_0$ vary with the number of nodes of a network and the assumed ranges of vector (host) recruitment rates with ranges of other parameters fixed. The values of $R_0$  are sorted from small values to large valued in Figure $\ref{fig:SIR0<1}$  and $\ref{fig:SIR0>1}$, and $E_0$ are ranked from small values to large values in  Figure $\ref{fig:SIrhoM<1}$ and $\ref{fig:SIrhoM>1}$. The largest value of $E_0$  is $0.9980$ when all values of $R_0$ are smaller than $1$ and $R_0 \leq E_0$, as  shown in Figure $\ref{fig:SIR0<1}$. The smallest value of $E_0$  is $1.003$ when all values of $R_0$ are greater than $1$ and $R_0 \geq E_0$, as shown in Figure
$\ref{fig:SIR0>1}$. The largest value of $R_0$  is $0.9947$ when all values of $E_0$  are smaller than $1$, as  shown in Figure $\ref{fig:SIrhoM<1}$.  The smallest value of $R_0$  is $1.006$ when all values of $E_0$ are greater than $1$, as shown in Figure $\ref{fig:SIrhoM>1}$.
The value of  $E_0$ is not  monotonically increasing with the increase of $R_0$, as shown  Figure $\ref{fig:SIR0<1}$ and $\ref{fig:SIR0>1}$. Similarly,   $R_0$ fluctuates as $E_0$ increases, as shown in Figure $\ref{fig:SIrhoM<1}$ and $\ref{fig:SIrhoM>1}$.

\begin{figure}[!h]
\centering
\subfigure[When $ \sqrt{R_0} \leq 1 $,  $\sqrt{R_0} \leq  E_0$.]{
\label{fig:SIRR0<1}
\includegraphics[angle=0,width=6.1cm,height=5.9cm]{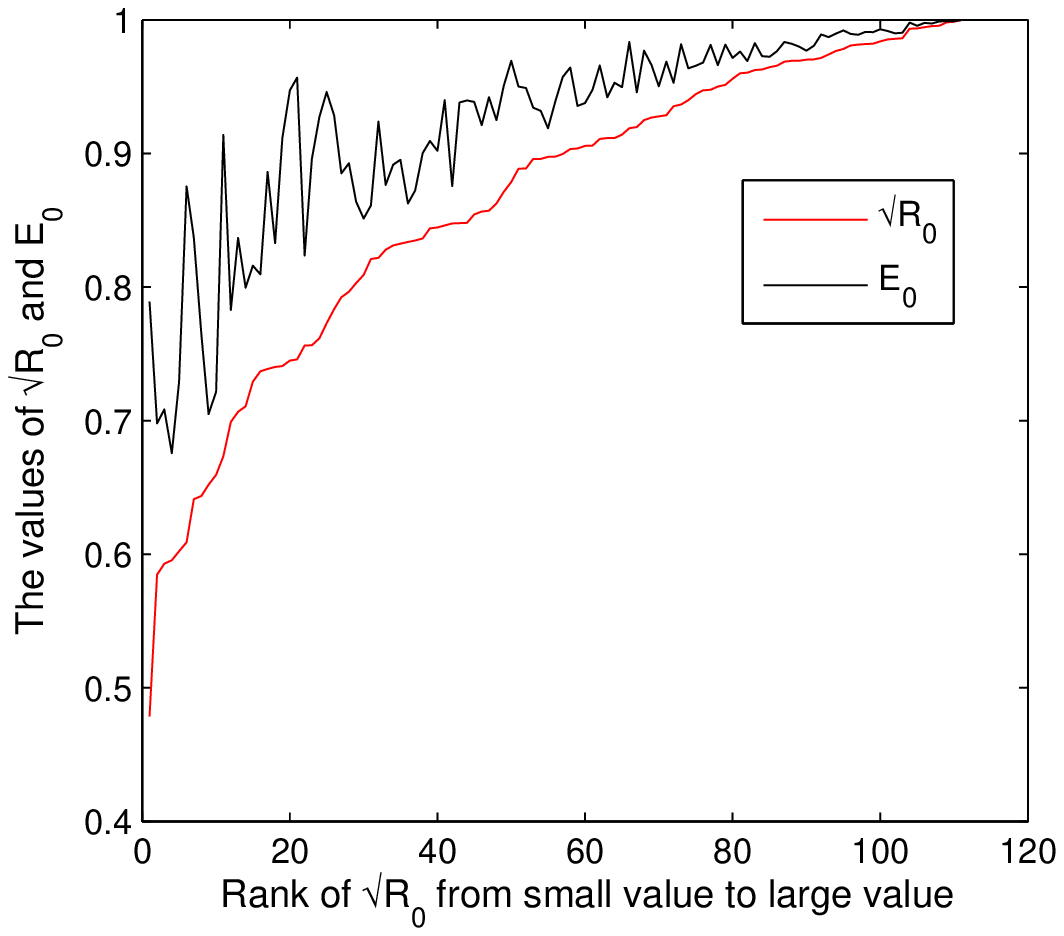}}
\hspace{0.1in}
\subfigure[When $\sqrt{R_0} \geq 1 $,  $\sqrt{R_0}  \geq  E_0$.]{
\label{fig:SIRR0>1}
\includegraphics[angle=0,width=6.1cm,height=5.9cm]{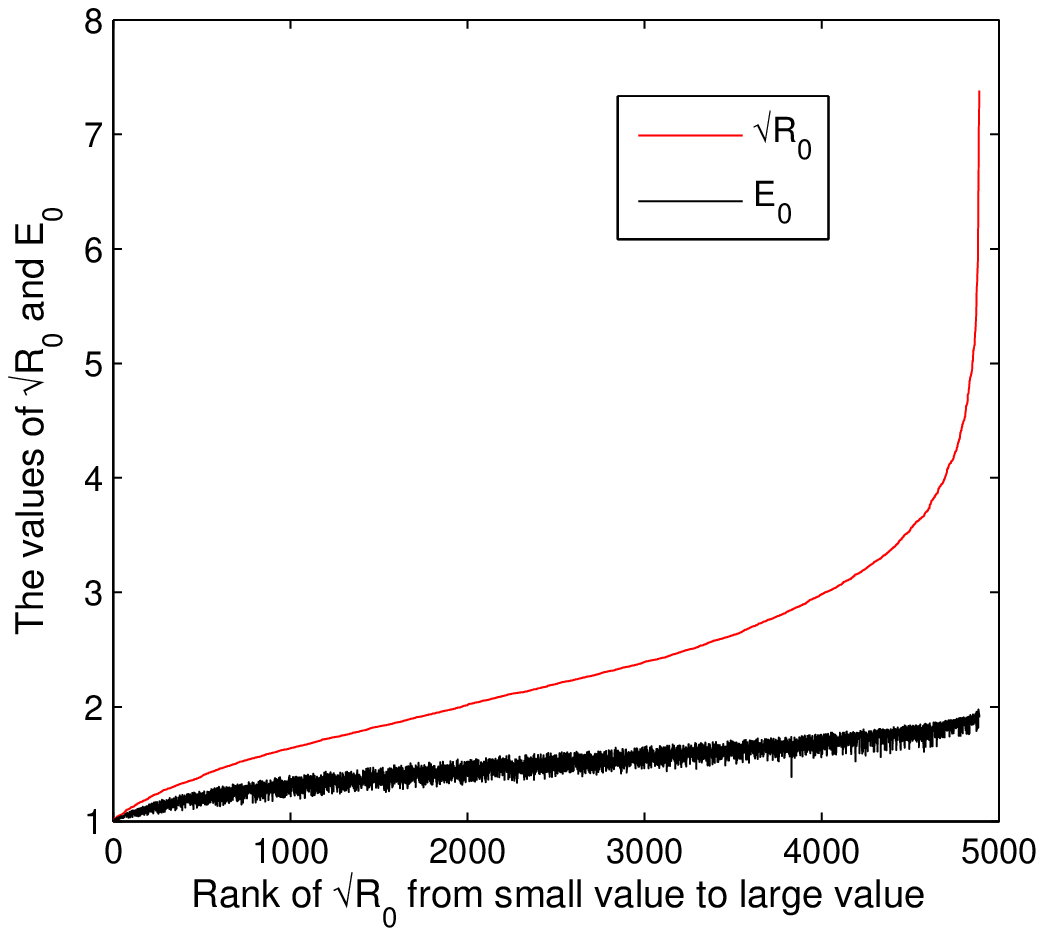}}
\hspace{0.1in}
\subfigure[When $E_0 \leq 1 $,  $\sqrt{R_0} \leq  E_0$.]{
\label{fig:SIRrhoM<1}
\includegraphics[angle=0,width=6.1cm,height=5.9cm]{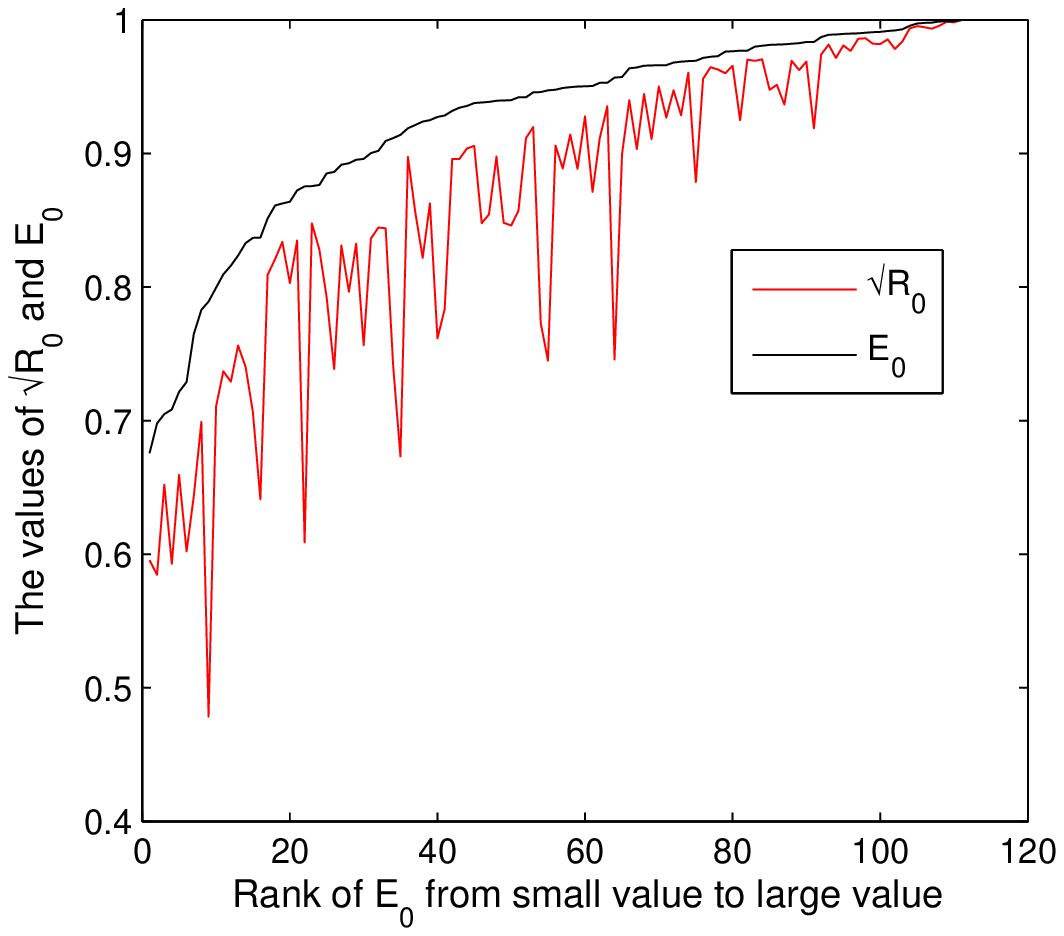}}
\hspace{0.1in}
\subfigure[When $E_0 \geq 1 $,  $\sqrt{R_0} \geq  E_0$.]{
\label{fig:SIRrhoM>1}
\includegraphics[angle=0,width=6.1cm,height=5.9cm]{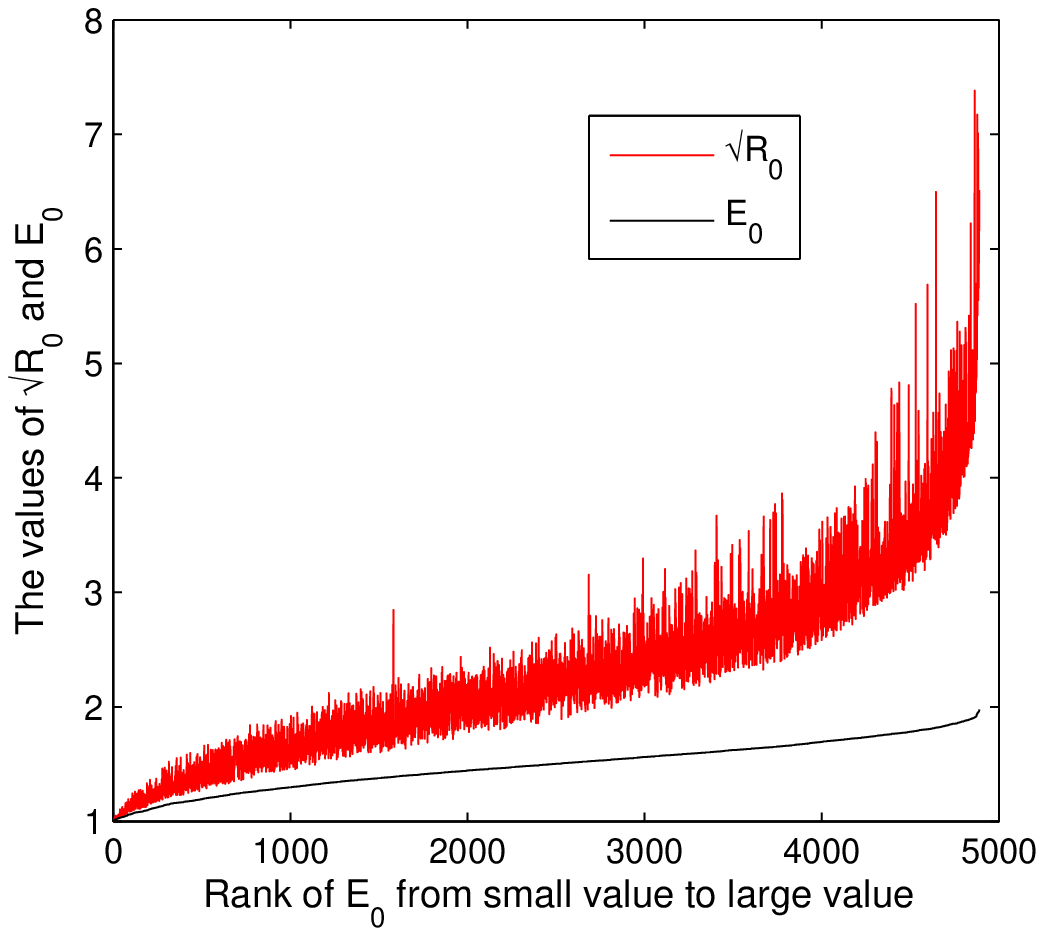}}
\hspace{0.1in}
\caption{Relationships between $R_0$ and $E_0$ for Rift Valley fever model.}
\label{fig:RVFmodel}
\end{figure}

\begin{table}
\centering
\begin{tabular}{|p{50pt}|p{130pt}|p{80pt}|p{45pt}|p{45pt}|}
 \hline
Parameter & Description & Range& Dimension & Source\\
\hline
$\alpha$  &  Contact rate: mosquitoes to humans & $0.010-0.27$& $1/$day &\cite{Chitnis2008}\\
\hline
$\beta$ &Contact rate: humans to mosquitoes  & $0.072-0.64$ &$1/$day &\cite{Chitnis2008}\\
\hline
$\mu$ &  Per capita death rate for mosquitoes \rm  &$0.020-0.27$ &$1/$day &\cite{Chitnis2008}\\
\hline
$d$ & Per capita death rate for humans  &$ 0.000027-0.00014$  & $1/$day &\cite{Chitnis2008}\\
\hline
$\gamma$ & Per capita recovery rate for humans   & $0.0014-0.0017$   &$1/$day & \cite{Chitnis2008}\\
\hline
$\eta$&  Mosquito recruitment  rate   &$1-5$   &$1/$day &Assume\\
\hline
$\psi$&  Human recruitment rate &$1-60$   &$1/$day &Assume\\
\hline
\end{tabular}
\caption{Parameters of the malaria metapopulation model.}
\label{table:malariaparameters}
\end{table}
Model $(\ref{SEIRmodel})$ is applied to numerical examine the relationship between $R_0$ and $E_0$ for  Rift Valley fever. See Table \ref{table:RVFparameters} for descriptions and ranges of parameters.  Five thousand realizations  produce $R_0$ ranging between $0.2289$ and  $54.5086$ and $E_0$ from $0.6757$ to $1.9763$.  The values of $R_0$ are ordered from small to large magnitudes in Figure $\ref{fig:SIRR0<1}$ and $\ref{fig:SIRR0>1}$, and the values of $E_0$ are from small to large values in    Figure $\ref{fig:SIRrhoM<1}$ and $\ref{fig:SIRrhoM>1}$.   The largest value of $E_0$  is $1$ when all values of $R_0$ are smaller than $1$,  and $\sqrt{R_0} \leq E_0$, as shown in Figure $\ref{fig:SIRR0<1}$. The smallest value of $E_0$  is $1.005$ when all values of $R_0$ are greater than $1$, and $\sqrt{R_0} \geq E_0$, as  shown in Figure
$\ref{fig:SIRR0>1}$. The largest value of $R_0$  is $0.9998$ when all values of $E_0$  are smaller than $1$, and $\sqrt{R_0} \geq E_0$,  as  shown in Figure $\ref{fig:SIRrhoM<1}$.  The smallest value of $R_0$  is $1.008$ when all values of $E_0$ are greater than $1$, and $\sqrt{R_0} \geq E_0$, as  shown in Figure $\ref{fig:SIRrhoM>1}$.
When $R_0$ increases, $E_0$ does not always increase, as  shown in Figure $\ref{fig:SIRR0<1}$ and $\ref{fig:SIRR0>1}$. Similarly,   $R_0$ fluctuates as $E_0$ increases, as  shown in Figure $\ref{fig:SIRrhoM<1}$ and $\ref{fig:SIRrhoM>1}$.

\begin{table}
\centering
\begin{tabular}{|p{50pt}|p{130pt}|p{80pt}|p{45pt}|p{62pt}|}
 \hline
Parameter & Description & Range& Dimension & Source\\
\hline
$\alpha$ &Contact rate: mosquito to  livestock   & $0.0021- 0.2762$ &$1/$day &\cite{Canyon1999,Hayes1973,Jones1985,Magnarelli1977,PrattMoore1993,Turell1988,Turell1988b}\\
\hline
$\beta$  &  Contact rate:   livestock to mosquitoes & $0- 0.32$& $1/$day &\cite{Canyon1999,Hayes1973,Jones1985,Magnarelli1977,PrattMoore1993,Turell1987}\\
\hline
$1/\mu$ &   Longevity of \ mosquitoes                        &$ 3-60$            &$1/$day &\cite{Bates1970,Moore1993,PrattMoore1993}\\
\hline
$1/d$ & Longevity of livestock                                        &$360-3600$          &$1/$day &\cite{Radostits2001}\\
\hline
$1/\gamma$& Recover rate in livestock                  &$2-5$  &$1/$day &\cite{Erasmus1981}\\
\hline
$1/\varphi$ &Incubation period in  mosquitoes &$4-8$  &days      &\cite{Turell1988}\\
\hline
$1/\epsilon$ &Incubation period in livestock                &$2-6  $   &days      &\cite{Peters1994}\\
\hline
$\eta$&  Mosquito recruitment  rate   &$1-500$   &$1/$day &Assume\\
\hline
$\psi$& Livestock recruitment rate &$1-10$   &$1/$day &Assume\\
\hline
\end{tabular}
\caption{Parameters of the Rift Valley fever metapopulation model.}
\label{table:RVFparameters}
\end{table}

\subsection{Sensitivity analysis}
We employ  Latin Hypercube Sampling/Partial Rank Correlation Coefficient (LHS/PRCC) analysis \cite{Marino2008}  to identify key parameters whose uncertainty contribute to predict uncertainty of $E_0$ for model $(\ref{SEIRmodel})$ and rank the parameters by their significances.  The parameters shown to be significant with large PRCC values $(>0.5)$ or small $p$-values $(<0.05)$ \cite{Gomero2012} by the sensitivity test with $5000$ sets of parameter values, are listed in Table \ref{table:EXT2013G16PRCC}. The magnitude of PRCC value represent the contribution to the prediction for the imprecision of $E_0$, and a negative sign indicates that the parameter is inversely proportional to the magnitude of $E_0$. The closer PRCC value is to $+1$ or $-1$, the more greatly the parameter  impacts the outcome of $E_0$.
\begin{table}
\centering
\begin{tabular}{|c| c|c|}
 \hline
Parameter & PRCC  & p-value\\
\hline
$\alpha$& $0.5649$&$<0.001$\\
\hline
$\beta$& $0.6039$&$<0.001$\\
\hline
$1/\mu$ &$0.8061$ &$<0.001$\\
\hline
$1/d$ &$-0.4660$ &$<0.001$\\
\hline
$1/\gamma$ &$0.5524$ &$<0.001$\\
\hline
$1/\varphi$   &$-0.0284$ &$<0.05$\\
\hline
$\eta$ &$0.5785$ &$<0.001$\\
\hline
$\psi$ &$-0.5036$ &$<0.001$\\
\hline
\end{tabular}
\caption{Sensitivity testing results using  Partial Rank Correlation Coefficients  for model $(\ref{SEIRmodel})$ for homogeneous populations. Only significant parameters are shown.}
\label{table:EXT2013G16PRCC}
\end{table}
\subsection{Trends of extinction array with varying parameters}
Consistent  trends of $w^*$ are observed  by numerical simulations for  homogeneous populations and a two-node network for Model $(\ref{SEIRmodel})$. Table $\ref{table:EXT2013G19}$ lists three different values for each parameter and corresponding extinction array    for  homogeneous populations as an example.  Table $\ref{table:SEIRparameterschange}$ shows the trends of extinction array  by varying one parameter at a  time, keeping other parameters fixed and $E_0>1$ for homogeneous populations and a two-node network. If at least one entry of extinction array   increases and others remain constant, then we say that the array increases. The extinction array $w^*$ decreases with the increase of  contact rates from local vectors and vectors in other nodes to local hosts, contact rates from local hosts and hosts in other nodes to local vectors, death rates of hosts, recruitment rates of vectors, and incubation rates of vectors and hosts,  whereas increases with the increase of vector death rates, vector recovery rates, host recruitment rates.

\begin{table}
\centering
\begin{tabular}{|c| c|}
 \hline
Changing parameter &  $(w_1^*, w_2^*,  u_1^*, u_2^*)$ \\
\hline
$\alpha=0.0601$  & $(0.9965, 0.9978, 0.9961, 0.9978)$  \\
\hline
$\alpha=0.0766$ & $(0.8648, 0.9212, 0.8467, 0.9212)$\\
\hline
$\alpha=0.0781$  & $(0.8546, 0.9158, 0.8352, 0.9158)$ \\
\hline
$\beta=0.0639$  & $(0.9158, 0.9824, 0.9046, 0.9824)$  \\
\hline
$\beta=0.1026$ & $(0.6623, 0.8967, 0.6173, 0.8966)$\\
\hline
$\beta=0.1426$  & $(0.5448, 0.8224, 0.4841, 0.8223)$ \\
\hline
$\mu=1/60$  & $(0.1955, 0.4961, 0.1419, 0.4956)$  \\
\hline
$\mu=1/59$ & $(0.1996, 0.5016, 0.1453, 0.5110)$\\
\hline
$\mu=1/56$  & $(0.2127, 0.5188, 0.1565, 0.5182)$ \\
\hline
$d=1/3477$  & $(0.4621, 0.7398, 0.3904, 0.7395)$  \\
\hline
$d=1/3370$ & $(0.4554, 0.7312, 0.3828, 0.7310)$\\
\hline
$d=1/3311$  & $(0.4518, 0.7265, 0.3787, 0.7262)$ \\
\hline
$\gamma=1/5$   & $(0.4247, 0.6877, 0.3480, 0.6874)$  \\
\hline
$\gamma=1/4$  & $(0.4698, 0.7491, 0.3992, 0.7488)$\\
\hline
$\gamma=1/3$ & $(0.5451, 0.8226,  0.4845, 0.8224)$ \\
\hline
$\epsilon=1/6$  & $(0.4700, 0.7493, 0.3994, 0.7489)$  \\
\hline
$\epsilon=1/4$ & $(0.4698, 0.7491, 0.3992, 0.7488)$\\
\hline
$\epsilon=1/2$  & $(0.4697, 0.7489, 0.3990, 0.7488)$ \\
\hline
$\varphi=1/8$   & $(0.5494, 0.7784, 0.4293, 0.7782)$  \\
\hline
$\varphi=1/7$  & $(0.5312, 0.7715, 0.4218, 0.7712)$\\
\hline
$\varphi=1/6$ & $(0.5119, 0.7643,  0.4142, 0.7641)$ \\
\hline
$\eta=19$   & $(0.5412, 0.8195, 0.4801, 0.8193)$  \\
\hline
$\eta=76$  & $(0.5264, 0.8069, 0.4632, 0.8066)$\\
\hline
$\eta=482$ & $(0.2907, 0.3169,  0.1961, 0.3162)$ \\
\hline
$\psi=1$   & $(0.4698, 0.7491, 0.3992, 0.7488)$  \\
\hline
$\psi=2$  & $(0.6859, 0.9123, 0.6553, 0.9122)$\\
\hline
$\psi=3$ & $(0.9219, 0.9838,  0.9115, 0.9838)$ \\
\hline
\end{tabular}
\caption{The extinction array   changes  with one parameter within the range at a time for  homogeneous populations, while keeping other parameters fixed and $E_0>1$ for model $(\ref{SEIRmodel})$. Fixed parameters are: $\alpha=0.2$, $\beta=0.19$, $\mu=1/30$, $d=1/3600$, $\gamma=1/4$,  $\epsilon=1/2$,  $\varphi=1/4$,  $\eta=100$, $\psi=1$ in this example. Same trends are obtained with various sets of fixed parameters. }
\label{table:EXT2013G19}
\end{table}

\begin{table}
\centering
\begin{tabular}{|c| c|}
 \hline
Increasing parameter &  $(w_1^*,  \cdots, w_n^*, u_1^*, \cdots, u_n^*)$ \\
\hline
$\alpha_i$,  $\beta_i$,   $d_i$,  $\varepsilon_i$, $\varphi_i$,  $\eta_i$,  $\sigma_{ij}$, $\omega_{ij}$   $(i, j=1,  \cdots, n, i \neq j )$& decreases \\
\hline
$\mu_i$, $\gamma_i$,  $\psi_i$   $(i=1,  \cdots, n)$& increases \\
\hline
\end{tabular}
\caption{Summary of trends for  extinction array changing with one parameter  at a time, while keeping other parameters fixed and $E_0>1$ for model $(\ref{SEIRmodel})$  for homogeneous populations  and a two-node network throughout various simulations.}
\label{table:SEIRparameterschange}
\end{table}

\section{Discussions}
\label{section:result}
The reproduction number, $R_0$ for deterministic vector-host models and  thresholds for extinction probabilities, $E_0$  for corresponding CTMC models are analytically and numerically connected. For  model $(\ref{equation:SImodel})$, our analysis show that $R_0\leq 1$, if and only if $E_0\leq 1$,  and  $|R_0-1|\geq |E_0-1|$  under certain assumptions. 
 Numerical simulations for a malaria model   on heterogeneous networks with different number of nodes  show that Corollary $\ref{cor1}$ holds without any assumptions. For  model $(\ref{SEIRmodel})$,  analytical  results show that  $\sqrt{R_0}<1$ if and only if $E_0<1$, and $|\sqrt{R_0}-1|\geq |E_0-1|$  by the same assumption in $(\ref{eq-assumption})$.  Extensive numerical simulation results  for a Rift Valley fever model on networks with various number of nodes show that
 Corollary $\ref{cor2}$  holds without any assumptions.


\begin{conjecture}
Theorems $\ref{cor1}$, $\ref{cor2}$ and Corollary  $\ref{cor1}$,  $\ref{cor2}$  hold without  assumption $(\ref{eq-assumption})$, i.e.,  $R_0 \leq 1$  if and only if   $E_0\leq 1$  for both models $(\ref{equation:SImodel})$ and $(\ref{SEIRmodel})$, besides, $|R_0-1|\geq |E_0-1|$ for model $(\ref{equation:SImodel})$, and $|\sqrt{R_0}-1|\geq |E_0-1|$  for model $(\ref{SEIRmodel})$ without assumption $(\ref{eq-assumption})$.
\end{conjecture}

The first part was proven  by Allen and van den Driessche under the assumption $(16)$ in \cite{ Lindarelation2013}, i.e., $(F-V)^T=W(M-I)$, where $F$ and $V$ are Jacobian matrices defined in $(\ref{Jacobian})$, $M$ is a mean matrix of offspring distribution  defined in Section $\ref{sec:threshold}$,
$I$ is  the identity matrix, and $W$ is a positive diagonal matrix with each entry $w_i$ representing the rate parameter at which  lifespan of  group $i$  are exponentially distributed for $ i=1, \cdots n$  \cite{Penisson2010}. This assumption does not hold for both model $(\ref{equation:SImodel})$ and model $(\ref{SEIRmodel})$.

Consistent trends in the extinction array  $w^*$  while changing one parameter through numerical simulations is helpful in deducing  trends of extinction probability and possible interventions for vector-borne diseases. According to  Equation   $(\ref{extprop})$, the probability of disease extinction  is monotonically increasing (decreasing) with  the increase (decrease) of  the extinction array   when the initial number of infection is fixed. The following  biological interpretations on disease extinction or persistence are in terms of fixed initial number of infections.  If contact rates from vectors to hosts $(\alpha, \sigma)$, or those from hosts to vectors  ($\beta, \omega$) increase, the probability for the disease to persist is higher. If death rates of hosts $(d)$ increase, the number of vectors is relatively dominant. Consequently, the disease is more likely to persist. Similarly, growing vector recruitment rates ($\eta$) increase the probability for disease persistence. The higher incubation rates in vectors ($\varphi$) or that in hosts ($\epsilon$)  lead to faster vector or  host infections, such  that the disease  is prone to persist. On the contrary, increasing death rates of vectors ($\mu$)  may reduce rates of  host infection, and ultimately,  may increase the likelihood of disease extinction. Increasing recovery rates of hosts ($\gamma$) may reduce the number of infections, such that  the probability of disease extinction increases. Increasing recruitment rate of hosts ($\psi$) may reduce vector infection rates and increase probability of disease extinction.

The findings show that we may increase extinction probability of vector-borne diseases by properly controlling vector and host population size, and promptly detecting and applying treatment for hosts. Analytical  and numerical results shed light on   deriving relationships between $R_0$ and $E_0$, as well as connections between varying parameters  and increasing extinction probabilities  for many other vector-borne diseases transmitted among heterogeneous works.
In summary, the resulting  mathematical derivations and numerical simulations facilitate understanding  thresholds for the spread of vector-borne diseases, as well as providing novel  insights into disease prevention, mitigation and control strategies.

\section*{Acknowledgments}
This material is based upon work supported by the U. S. Department of Homeland Security under Grant Award Number 2010-ST061-AG0001, a grant through the Kansas Biosciences Authority, and is funded by U. S.  Department of Agriculture - the Agricultural Research Service with Specific Cooperative Agreement 58-5430-1-0356. The views and conclusions contained in this publication are those of the authors and should not be interpreted as necessarily representing the official policies, either explicit or implicit, of the U. S. Department of Homeland Security and Kansas Biosciences Authority.  We are grateful to Dr. Lee Cohnstaedt and Dr. H. Morgan Scott, and Dr. Linda J. S. Allen for helpful conversations.

\bibliographystyle{spmpsci}
\bibliography{extinctionJMB1}
\end{document}